\documentstyle[12pt]{article}
\input tcilatex
\QQQ{Language}{
American English
}

\begin{document}

DEDUCTION OF THE QUANTUM NUMBERS OF LOW-LYING STATES

OF THE ($e^{+}e^{+}e^{-}e^{-}$) SYSTEM FROM SYMMETRY CONSIDERATION

\hspace*{1.0in}

C.G.Bao

Department of Physics, Zhongshan University, Guangzhou, 510275, P.R.China

\hspace*{1.0in}

ABSTRACTS The features of the low-lying spectrum and a complete set of
quantum numbers for the ($e^{+}e^{+}e^{-}e^{-}$) system have been deduced
based on symmetry. The existence of a low odd-parity L=1 excited state with
the two $e^{-}$ coupled to s$_1=0$, and two $e^{+}$ coupled to s$_2=1$ (or s$%
_1$ =1 and s$_2$ =0) and a low even-parity L=0 excited state with s$_1=s_2=1 
$ have been anticipated. A 2-dimensional ($e^{+}e^{+}e^{-}e^{-}$) system has
also been discussed.

\hspace{1.0in}


\hspace{1.0in}


\newpage\ 

The systems composed of a few electrons and positrons are known to exist in
nature. The existence of the ($e^{+}e^{-}$) with infinite bound states is
obvious. The ($e^{+}e^{-}e^{-}$) or ($e^{+}e^{+}e^{-}$) has been
investigated by many authors. In 1977 Hill proved that these two 3-body
systems are bound [1]. The ground states and a number of resonances have
been identified ([2,3] and references there in). For the ($%
e^{+}e^{+}e^{-}e^{-}$) a direct experimental observation has not yet been
reported. However, a few theoretical calculations [4-8] indicate that this
system is bound. A recent (and more accurate) calculation reports that the
ground state has an energy E$_0$ = -0.515989 (atomic units are used in this
paper) and a radius r$_0$ =3.608 [8]. From r$_0$ one can estimate the moment
of inertia I$_0$ of the ground state. It was found that the $\frac{\hbar ^2}{%
2\text{I}_0}$ is only about $\frac 1{50}\left| \text{E}_0\right| $. Since
for Coulombic systems the excited states would be in general much larger
than the ground state in size (as shown in [6]), it implies that an
excitation of collective rotation is very easy to realize in an excited
state due to having a large moment of inertia. . Therefore, besides the
orbital angular momentum L=0 resonances predicted before [5,6], the
existence of a number of resonances with L$\neq $ 0 is very possible. Thus,
the spectrum of the ($e^{+}e^{+}e^{-}e^{-}$) would be rich. However, a
precise calculation is difficult. Before solving the Schr\"odinger equation,
if the qualitative feature can be more or less understood, it would be very
helpful to understand the coming results from experiments and theoretical
calculations. It is believed that symmetry will play an essential role in
the spectrum. Based on the permutation symmetry, Kinghorn and Poshusta have
proposed a primary scheme for the classification of states [6]. On the other
hand, the feature of a quantum state depends on how the wave function is
distributed in the coordinate space. It turns out that the distribution is
decisively affected by symmetry (not only by the permutation symmetry, as we
shall see). In this paper the effect of symmetry has been investigated in
detail, thereby the geometric feature of low-lying states has been
extracted, a complete set of quantum numbers (including those proposed in
[6]) has been introduced for the classification.

Let H be the internal Hamiltonian for the ($e^{+}e^{+}e^{-}e^{-}$). Let
particles 1 and 2 be electrons, 3 and 4 be positrons. It was pointed out by
Kinghorn and Poshusta that H is invariant under the permutations

((e)), ((12)), ((34)), ((12))((34)), p(1324), p(1423), ((13))((24)), and
((14))((23)),

here the ((12)) denotes an interchange of 1 and 2, p(1324) denotes a cyclic
permutation, etc.. These permutations form a subgroup of the S$_4$ group
isomorphic to the point group D$_{2d}$ [6]. Therefore the eigenstates of H
can be classified according to the representations $\mu $ of the D$_{2d}$
group, $\mu =$ A$_1$ , A$_2$, B$_1$, B$_2$, or E. These representations can
be induced by the following idempotents:

e$^{A_1}=\frac 18[1+((13))((24))][1+((12))][1+((34))]$\hspace{1.0in}(1,1)

e$^{A_2}=\frac 18[1-((13))((24))][1-((12))][1-((34))]$\hspace{1.0in}(1,2)

e$^{B_1}=\frac 18[1+((13))((24))][1-((12))][1-((34))]$\hspace{1.0in}(1,3)

e$^{B_2}=\frac 18[1-((13))((24))][1+((12))][1+((34))]$\hspace{1.0in}(1,4)

e$^{E_{11}}=\frac 14[1-((12))((34))+((13))((24))-((14))((23))]$\hspace{1.0in}%
(1,5)

e$^{E_{22}}=\frac 14[1-((12))((34))-((13))((24))+((14))((23))]$\hspace{1.0in}%
(1,6)

Evidently, the sum of these six idempotents is equal to one.

On the other hand, H is also invariant under rotation and space inversion.
Besides, the total spin s$_1$ of particles 1 and 2 and the total spin s$_2$
of 3 and 4 are also good quantum numbers. Let the spatial wavefunctions of
the eigenstates be denoted as $\Psi .$ Evidently, $\Psi $ should be further
classified according to the total orbital angular momentum L, parity $\Pi $,
s$_1$ and s$_2$. In particular, we have

((12))$\Psi =(-1)^{s_1}\Psi $\hspace{1.0in}(2,1)

and

((34))$\Psi =(-1)^{s_2}\Psi $\hspace{1.0in}(2,2)

Equations (2,1) and (2,2) are additional conditions imposed on the
eigenstates of H.

For an eigenstate $\Psi $ belonging to the A$_1$-representation, it is
evident from eq.(1,1) that

((12))$\Psi =\Psi ,$ ((34))$\Psi =\Psi ,$ and ((13))((24))$\Psi =\Psi $

Therefore this state has s$_1$=0, s$_2$=0, and $\Lambda $=1, where $\Lambda $
is the eigenvalue of ((13))((24)) . Similarly, the states belonging to the A$%
_2$-representation have s$_1$=1, s$_2$=1, and $\Lambda =-$1; the states of B$%
_1$ have s$_1$=1, s$_2$=1, and $\Lambda $=1; and the states of B$_2$ have s$%
_1$=0, s$_2$=0, and $\Lambda $=$-$1 (cf. Table 1).

In the case of the E-representation, there are two degenerate states for an
eigenenergy. They can be denoted as e$^{E_{11}}\Psi $ and ((12))e$%
^{E_{11}}\Psi $ (or e$^{E_{22}}\Psi $ and ((12))e$^{E_{22}}\Psi $). However,
neither the e$^{E_{11}}\Psi $ nor the ((12)) e$^{E_{11}}\Psi $ fulfills
(2.1) and (2.2), therefore each of them is not a physical solution.
Nonetheless, the two linear combinations

$\Psi _a=[1-((12))]$e$^{E_{11}}\Psi $\hspace{1.0in}(3,1)

and

$\Psi _b=[1+((12))]$e$^{E_{11}}\Psi $\hspace{1.0in}(3,2)

are physical solutions. One can prove that ((12))$\Psi _a=-\Psi _a$, (($34)$)%
$\Psi _a=\Psi _a$, and ((13))((24))$\Psi _a=\Psi _b$. Therefore the $\Psi _a$
has s$_1$=1 and s$_2$=0, but it does not have the quantum number $\Lambda .$
For convenience, this state is denoted as $\Lambda $=0. Similarly, the $\Psi
_b$ has s$_1$=0, s$_2$=1, and $\Lambda $=0. The quantum numbers L,$\Pi ,$
and $\mu $ are sufficient to label the symmetry of a series of states. Thus
a state is labeled as L$_i^\Pi \{\mu \}$ in the following, where $i$ denotes
the ordering of the state in the (L$\Pi \mu )$ series. More explicitly, the
spatial wavefunction of a L$_i^\Pi \{\mu \}$ state is denoted in more detail
as $F_{M,s_1s_2}^{L\Pi \Lambda }(1234),$ $M$ is the Z-component of L. The
relation between $\mu $ and the set (s$_1$s$_2\Lambda $) is listed in Table
1.

.In order to facilitate the analysis, a body frame i'-j'-k' is introduced.
Then the wave function can be expanded as

\hspace{1.0in}

$F_{M,s_1s_2}^{L\Pi \Lambda }(1234)=\sum_Q D_{QM}^L(-\gamma
,-\beta ,-\alpha )F_{Q,s_1s_2}^{L\Pi \Lambda }(1^{\prime }2^{\prime
}3^{\prime }4^{\prime })$\hspace{1.0in}(4)

where $Q$ is the component of L\ along the third body axis k', $\alpha \beta
\gamma $ are the Euler angles, $D_{QM}^L$ is the Wigner function of
rotation. Here the (1234) and (1'2'3'4') specify that the coordinates are
relative to a fixed frame and to the body-frame, respectively. $%
F_{Q,s_1s_2}^{L\Pi \Lambda }(1^{\prime }2^{\prime }3^{\prime }4^{\prime })$
is called a Q-component. We shall show that the Q-components are seriously
constrained by symmetry.

Evidently, the distribution of the wave function in the coordinate space
depends first on the interactions. For the ($e^{+}e^{+}e^{-}e^{-}$) system,
if the particles form a shape where the repulsive force between the pair of
electrons (or positrons) can not be over compensated by attractive force,
then the shape is not important for low-lying states , and the amplitude of
wave function is expected to be small at this shape. This is the case, e.g.,
that the particles form a regular tetrahedron or an quadrilateral with the
pair of electrons (positrons) located at adjacent vertexes. On the other
hand, if the particles form an quadrilateral with the pair of electrons
(positrons) located at the two ends of a diagonal, then the configuration is
favorable to binding, where each repulsive bond can be over compensated.
When the particles form a square with the same kind of particles located at
the two ends of each diagonal, this particular configuration is labeled as
SQ hereafter.

Let the third body axis k' be normal to the plane of the SQ, we have

(i) a rotation of the SQ about k' by 180$^{\circ }$ is equivalent to a space
inversion , thus we have

\hspace{1.0in}

$(-1)^QF_{Q,s_1s_2}^{L\Pi \Lambda }(SQ)=\Pi F_{Q,s_1s_2}^{L\Pi \Lambda
}(SQ)$\hspace{1.0in} (5)

\hspace{1.0in}

here the $(SQ)$ implies that the coordinates form a SQ.

(ii) a rotation about k' by 180$^{\circ }$ is equivalent to ((12))((34)) ,
thus we have

\hspace{1.0in}

$(-1)^QF_{Q,s_1s_2}^{L\Pi \Lambda }(SQ)=(-1)^{s_1+s_2}F_{Q,s_1s_2}^{L\Pi
\Lambda }(SQ)$\hspace{1.0in} (6)

\hspace{1.0in}

(iii) Let i' be parallel to $\stackrel{\rightarrow }{r_{12}}$ lying along a
diagonal, a rotation about i' by 180$^{\circ }$ is equivalent to ((34)),
thus we have

\hspace{1.0in}

$(-1)^LF_{\stackrel{-}{Q},s_1s_2}^{L\Pi \Lambda
}(SQ)=(-1)^{s_2}F_{Q,s_1s_2}^{L\Pi \Lambda }(SQ)$\hspace{1.0in} (7)

\hspace{1.0in}

(iv) a rotation about k' by -90$^{\circ }$ is equivalent to $p(1324),$ thus
we have

\hspace{1.0in}

$i^QF_{Q,s_1s_2}^{L\Pi \Lambda }(SQ)=\Lambda (-1)^{s_2}F_{Q,s_1s_2}^{L\Pi
\Lambda }(SQ)\hspace{1.0in} ($ if $\Lambda =\pm 1)\hspace{1.0in}(8.1)$

\hspace{1.0in}

$i^QF_{Q,s_1s_2}^{L\Pi \Lambda }(SQ)=(-1)^{s_2}F_{Q,s_2s_1}^{L\Pi \Lambda
}(SQ)\hspace{1.0in} ($ if $\Lambda =0)\hspace{1.0in}(8.2)$

\hspace{1.0in}

These equations impose a very strong constraint on  $F_{Q,s_1s_2}^{L\Pi
\Lambda }(SQ)$. If they can not be fulfilled,  $F_{Q,s_1s_2}^{L\Pi \Lambda }$%
must be zero at the SQ. In this case, a nodal surface exists at the SQ to
prevent the Q-component $F_{Q,s_1s_2}^{L\Pi \Lambda }$ from accessing the
SQ. Evidently, this nodal surface arises purely from symmetry, it is called
an inherent nodal surface [9,10]. The existence and the locations of the
inherent nodal surfaces do not at all depend on dynamics, it depends only on
the transformation property of the wavefunction under symmetric operations (
rotation, space inversion, and particle permutation), this property is
determined by the quantum numbers L,$\Pi ,$ and $\mu .$ All the Q-components
of the L$\leq 2$ states allowed to access the SQ are listed in Tab.1, they
are called SQ-accessible components.

All the other regular shapes in the neighborhood of the SQ have a weaker
geometric symmetry. For example, when a diagonal of a SQ becomes longer or
shorter, the shape becomes to a diamond. At this shape the equations (5) to
(7) hold, but (8) does not hold. Therefore the wavefunction is less
constrained at the diamonds. Hence, once a wavefunction can access the SQ,
it can access a broad domain surrounding the SQ. Therefore the SQ-accessible
components are inherent nodeless in this broad domain, where the
distribution of wave function can be optimized to favor binding.

When the particles form a straight chain with each adjacent pair separated
equidistantly and with each adjacent pair containing opposite charges, the
configuration is denoted by CH. Evidently, at the CH, the repulsion can also
be over compensated by attraction. However, all the four attractive bonds at
the SQ act effectively while only three bonds at the CH do. Thus, the CH is
less favorable than the SQ, and the domain surrounding the CH is the second
important domain for the low-lying states.

Let k' be normal to and i' be parallel to the straight chain, let the
ordering of the particles along the chain is 1423. Then, at the CH, a
rotation about k' by 180$^{\circ }$ is equivalent to a space inversion and
also equivalent to $p_{13}p_{24}$, a rotation about i' by an arbitrary angle 
$\delta $ causes no changes. Thus we have

\hspace{1.0in}

$(-1)^QF_{Q,s_1s_2}^{L\Pi \Lambda }(CH)=\Pi F_{Q,s_1s_2}^{L\Pi \Lambda
}(CH)$\hspace{1.0in}(9)

\hspace{1.0in}

$(-1)^QF_{Q,s_1s_2}^{L\Pi \Lambda }(CH)=\Lambda F_{Q,s_1s_2}^{L\Pi \Lambda
}(CH)$\hspace{1.0in} (if $\Lambda =\pm 1$)\hspace{1.0in}(10.1)

\hspace{1.0in}

$(-1)^QF_{Q,s_1s_2}^{L\Pi \Lambda }(CH)=F_{Q,s_2s_1}^{L\Pi \Lambda
}(CH)$\hspace{1.0in} (if $\Lambda =0$)\hspace{1.0in}(10.2)

and

$(1-i\delta L_{i^{\prime }}+\frac 12(i\delta L_{i^{\prime }})^2+\cdot \cdot
\cdot \cdot )F_{Q,s_1s_2}^{L\Pi \Lambda }(CH)=F_{Q,s_1s_2}^{L\Pi \Lambda
}(CH)$\hspace{1.0in}(11)

\hspace{1.0in}

where $L_{i^{\prime }}$ is the projection of L along i'. Evidently, (11) is
fulfilled if

\hspace{1.0in}

$L_{i^{\prime }}F_{Q,s_1s_2}^{L\Pi \Lambda }(CH)=0$\hspace{1.0in}(for all Q)
\hspace{1.0in}(12)

(12) can be rewritten as

\hspace{1.0in}

$\sqrt{(L-Q)(L+Q+1)}F_{Q+1,s_1s_2}^{L\Pi \Lambda }(CH)+\sqrt{(L+Q)(L-Q+1)}%
F_{Q-1,s_1s_2}^{L\Pi \Lambda }(CH)$

$=0$ \hspace{1.0in} (for all Q)\hspace{1.0in}(13)

\hspace{1.0in}

From this set of homogeneous linear equation, the $F_{Q,s_1s_2}^{L\Pi
\Lambda }(CH)$ are not zero only if

$(-1)^Q=(-1)^L$\hspace{1.0in} (14)

From (9), (10), and (14) the CH-accessible components can be identified as
listed in Tab.1. Incidentally, for a state with L$\geq 2,$ the components
with different \TEXTsymbol{\vert}Q\TEXTsymbol{\vert} will be mixed up
according to (13).

Let the L$_1^\Pi \{\mu \}$ state, the lowest one of the (L$\Pi \mu $)
series, be called a first-state (the subscript $i$ will be omitted if $i=1$%
). It is well known that higher states in general contain more nodal
surfaces. Thus it is reasonable to assume that the first-states would prefer
to contain nodal surfaces as least as possible. Therefore they would be
dominated by the component(s) containing least inherent nodal surfaces. In
this paper the analysis is based on the inherent nodal structure in the two
most important domain, i.e., the domain surrounding the SQ and the domain
surrounding the CH. There is no a sharp border to separate these domains.
Owing to the long range character of the Coulomb force, the distribution of
the wave functions is anticipated to be broad, and the wave functions will
extend from one domain to another. Hence, the component which is inherent
nodeless in both the SQ and CH domains is superior. In Table 1 there are
four such superior components denoted as SQ,CH. Accordingly, there are four
states with L$\leq 2$ superior in binding. Each of them is dominated by a
superior component., where the wave function is anticipated to be mainly
distributed in the SQ-domain but extend to the CH-domain without nodal
surfaces. These states are called the SQ1-states as listed in Table 2.

There is a type of component being SQ-accessible but CH-inaccessible
(denoted as SQ in Table 1). For the 1$^{+}\{A_2\},$2$^{+}\{B_2\},$2$%
^{-}\{E\},$ this type of component is much better than the other components
of them. Thus these three first-states would be dominated by this type of
component, and they are called the SQ2-states as listed in Table 2. The
internal energy of a SQ2-first-state is higher than a SQ1-first-state
because the former contains an inherent nodal surface while the latter does
not. .

There is a type of component being SQ-inaccessible but CH-accessible. They
can be further classified into three kinds. The first kind is not only
CH-accessible but also rectangle-accessible (denoted by CH1 in Table1). The
states dominated by this kind of components are the 0$^{+}\{B_1\}$ and 2$%
^{+}\{B_1\},$ they are called the CH1-states as listed in Table 2. The
second kind denoted as CH2 in Table 1 is rectangle-accessible if and only if
the particles at the two ends of a diagonal have opposite charges
(otherwise,it is rectangle-inaccessible). The states dominated by this kind
of components are the 1$^{-}\{A_2\}$ and 1$^{-}\{B_2\},$ they are called the
CH2-states. The third kind denoted as CH3 in Table 1 is
rectangle-inaccessible (disregarding how the particles are located at the
vertexes). The states dominated by this kind of components are the 0$%
^{+}\{E\}$ and 2$^{+}\{E\},$ they are called the CH3-states. The
wavefunctions of the CH1-, CH2-, and CH3-first-states would be mainly
distributed around a straight chain. Among them the internal energy of the
CH1-first-states would be lower due to being rectangle-accessible, the
CH3-first states would be higher due to being rectangle-inaccessible (thus
more nodal surfaces are contained).

There is a type of components being both SQ- and CH-inaccessible but
diamond-accessible (denoted as D-S-D in Table 1). The 0$^{+}\{B_2\},1^{+}%
\{B_1\},$ and 2$^{+}\{A_2\}$ would be dominated by this type of component
because they do not have a better choice. In these states coplanar structure
(e.g., a diamond) would exist. However, the inherent nodal surface at the
SQ\ will induce a coplanar oscillation back and forth around the SQ,
therefore the energy of these states are high. They are called the D-S-D
states listed in Table 2

There is a type of components being not allowed to access any coplanar
structure, they also can not access the CH, but they can access a regular
tetrahedron (denoted as T-S-T in Table 1). The 0$^{-}\{B_1\},\
1^{+}\{E\},\cdot \cdot \cdot \cdot \cdot $ would be dominated by this type
of component, where the wavefunction is mainly distributed around a
tetrahedron. However, the inherent nodal surface at the coplanar structure \
will induce an oscillation so that the tetrahedron would be transformed to
another tetrahedron via a coplanar structure. The energy of these states are
also high. They are called the T-S-T states listed in Table 2.

There is also a type of components being not allowed to access all the above
mentioned shapes, namely the SQ, rectangle, diamond, CH, and the regular
tetrahedrons (denoted as a blank block in Table 1). This type is very
unfavorable in binding due to containing many nodal surfaces at the regular
shapes. However, the 0$^{+}\{A_2\},\ 0^{-}\{A_1\},\cdot \cdot \cdot \cdot
\cdot $states contain only this type of component. The energy of these
states must be very high. They are listed at the bottom of Table 2.

A complete classification of the first-states according to the inherent
nodal structure of their dominant component is given in Table 2. The L$\geq 3
$ first-states can be classified as well. It is noted that  in Table 2 the
states in an upper row contains fewer inherent nodal surface; the upper, the
fewer.  For the first-states with a given L, the ordering of their locations
in  Table 2 (in descending order) is anticipated to be the ordering of their
energies (in ascending order): the SQ1-states at the top are the lowest
which are essentially inherent nodeless, while the states at the bottom are
the highest which contain a number of inherent nodal surfaces at regular
shapes. Nonetheless, the SQ2- and CH1-states both contain essentially one
inherent nodal surface, therefore their ordering can not be  simply 
determined by symmetry. This is also the situation of the D-S-D and T-S-T
states. In the second-states and even higher states, in addition to inherent
nodal surfaces, more nodal surfaces arising from a pure dynamical reason
(not from symmetry) will be contained. We are not going to the details of
these states. Nonetheless, they will also be seriously affected by the
inherent nodal structure as listed in Table 1.

Incidentally, the collective rotation energy of a coplanar structure depends
on the orientation of the structure relative to L. When the third body axis
k' is normal to the plane of the coplanar structure, the plane would be
better normal to L in a Q-component with a larger \TEXTsymbol{\vert}Q%
\TEXTsymbol{\vert}. Thus, for a coplanar shape, the larger the \TEXTsymbol{%
\vert}Q\TEXTsymbol{\vert}, the larger the moment of inertia, and thereby the
smaller the rotation energy. For example, although both the 2$^{+}\{B_1\}$
and 2$^{+}\{A_1\}$ are SQ1-states with L=2, the collective rotation energy
of the former is smaller because the SQ structure would appear in 
\TEXTsymbol{\vert}Q\TEXTsymbol{\vert}=2 component (refer to Table 1). This
effect would affect the ordering of levels. How serious is the effect
remains to be clarified.

It is anticipated that the low-lying spectrum would be dominated by the
SQ1-, SQ2-, and CH1-states. In particular, the ground state would be the 0$%
^{+}\{A_1\}$. The candidate of the first excited state would be the
odd-parity 1$^{-}\{E\}$ and the 0$^{+}\{B_1\}.$ The 1$^{-}\{E\}$ is two-fold
degenerate because the (s$_1$s$_2$) has two choices (0,1) and (1,0). The
internal energy of the 0$^{+}\{B_1\}$ as a CH1-state is higher than the 1$%
^{-}\{E\}$ as a SQ1-state, however the 0$^{+}\{B_1\}$ does not contain
collective rotation energy, thus a compensation can be gained. It is a pity
that experimental and theoretical data are so few that most of the
predictions can not be checked at present. Nonetheless, owing to the work by
Kinghorn and Poshusta [6], the predictions on L=0 even-parity states can be
partially checked. The results from [6] are listed in Table 3 to be compared
with Table 2 and Table 1. In Table 3 the $0^{+}$\{A$_1$\} is the lowest just
as anticipated. Besides, the ordering of levels is just as anticipated;
i.e., the energies of the SQ1-state, CH1-state, CH3-state, D-S-D state, and
the state belonging to the last row of Table 2 are in ascending order. Thus,
in general, the spectrum from [6] supports the above analysis and the
classification.

In the above analysis the features of the wavefunctions has been
demonstrated. Obviously, all these features deduced from symmetry
consideration together with the suggested classification scheme need a
further check. In particular, although the analysis covers all the L$\leq 2$
states, it does not mean that all of them can be experimentally observed.
The observation depends on the width of the resonance, which is a topic not
yet touched in this paper.

The above discussion can be easily generalized to a 2-dimensional $%
(e^{+}e^{+}e^{-}e^{-})$ system. In this case $\Pi =(-1)^L$ holds always,
therefore the label L$_i$\{$\mu \}$ is sufficient to denote a state. The
accessibility of the regular shapes to the L$_i$\{$\mu $\} states is
summarized in Table 4, which leads to a classification of states as listed
in Table 5. The ground state is also a 0\{A$_1$\} state. There are three
lower excited states, namely the 1\{E\},0\{A$_2$\} and 0\{B$_1$\}). It is
noticeable that the 0\{A$_2$\} is now low. This point is very different from
the 3-dimensional system where the 0$^{+}\{A_2\}$ is very high . We can
further predict that the 2\{B$_1$\} is the lowest L=2 state.

In the case of a 2-dimensional biexcitons in a quantum well [11], the
effective mass of the hole is not necessary equal to the mass of the
particle, therefore the Hamiltonian is no longer invariant to the D$_{2d}$
group. In this case, the above discussion should be modified.

The procedure of analysis proposed in this paper is quite general for
investigating the qualitative feature of low-lying states. In fact it can be
generalized to investigate  different kinds of systems with different
numbers of particles [12-13]. In any case, one has to clarify which domains
in the coordinate space are important (it depends on the interactions) to
the low-lying states, and to clarify the inherent nodal structure (it
depends purely on symmetry) in these important domains. The inherent nodal
structure would provide a solid frame to classify the states. Based on the
inherent nodal structure, one can understand not only the behavior of an
individual system, but also the systematics of spectra and the similarity of
different systems.

\hspace{1.0in}

ACKNOWLEDGMENT This work is supported by the NNSFC of the PRC, and by a fund
from the National Educational Committee of the PRC.

\hspace{1.0in}

REFERENCES

\hspace{1.0in}

[1] R.N.Hill, J. Math. Phys. 18 (1977) 2316

[2]A.M.Frolov, Zh. Eksp. Teor. Fiz. 92 (1987)1959

[3] Z. Chen and C.D.Lin, Phys. Rev. A42 (1990) 18

[4] M.A.Lee, P.Vashista, and R.K.Kalia, Phys. Rev. Lett. 51 (1983) 2422

[5] Y.K.Ho, Phys. Rev. A33 (1986) 3584; A39 (1989) 2709

[6] D.B.Kinghorn and R.D.Poshusta, Phys. Rev. A47 (1993) 3671

[7] P.M.Kozlowski and L.Adamowicz, Phys. Rev., A48 (1993) 1903

[8] K.Vagra and Y.Suzuki, Phys. Rev. C52 (1995) 2885

[9] C.G.Bao Few-Body Systems, 13 (1992) 41; Phys. Rev. A47 (1993) 1752

[10] W.Y.Ruan and C.G.Bao, Few-Body Systems, 14 (1993) 25

[11] J.Singh, D. Birkedal, V.G.Lyssenko, and J.M. Hvam, Phys. Rev. B53
(1996) 15909.

[12] C.G.Bao, X.Z.Yang and C.D.Lin, Phys. Rev. A55 (1997) 4168

[13] C.G.Bao, Chin. Phys. Lett., 14 (1997) 20; Commun. Theor. Phys. 28
(1997) 363; Phys. Rev. Lett. 79 (1997) 3475.

\newpage\ 

\begin{tabular}{|c|c|c|c||c||c|c||c|c|c|}
\hline
- &  &  &  & L=0 & L=1 & L=1 & L=2 & L=2 & L=2 \\ \hline
$\Pi $ & s$_1$s$_2$ & $\Lambda $ & $\mu $ & Q=0 & Q=0 & \TEXTsymbol{\vert}Q%
\TEXTsymbol{\vert}=1 & Q=0 & \TEXTsymbol{\vert}Q\TEXTsymbol{\vert}=1 & 
\TEXTsymbol{\vert}Q\TEXTsymbol{\vert}=2 \\ \hline
+ & 00 & 1 & A$_1$ & SQ, CH1 &  &  & SQ, CH1 &  & CH, D-S-D \\ \hline
+ & 00 & -1 & B$_2$ & D-S-D &  &  & D-S-D &  & SQ \\ \hline
+ & 11 & 1 & B$_1$ & CH1 & D-S-D &  & CH1 &  & SQ, CH1 \\ \hline
+ & 11 & -1 & A$_2$ &  & SQ &  &  &  & D-S-D \\ \hline
+ & 01, 10 & 0 & E & CH3 &  & T-S-T & CH3 & T-S-T & CH2 \\ \hline
- & 00 & 1 & A$_1$ &  &  &  &  &  & T-S-T \\ \hline
- & 00 & -1 & B$_2$ &  & T-S-T & CH2 &  &  &  \\ \hline
- & 11 & 1 & B$_1$ & T-S-T &  &  & T-S-T &  &  \\ \hline
- & 11 & -1 & A$_2$ &  &  & CH2 &  &  & T-S-T \\ \hline
- & 01, 10 & 0 & E &  &  & SQ, CH1 &  & SQ &  \\ \hline
\end{tabular}

Table 1, The accessibility of the regular shapes to the Q-components of the L%
$^\Pi \{\mu \}$ states. A block with a SQ (CH) denotes that the
corresponding component can access the SQ (CH). If both the SQ and CH are
contained in a block, both can be accessed.

\hspace*{1.0in}

\hspace{1.0in}

\begin{tabular}{|cc|}
\hline
\multicolumn{1}{|c|}{SQ1-states} & 0$^{+}\{A_1\},$1$^{-}\{E\},$2$^{+}\{B_1\},
$2$^{+}\{A_1\}$ \\ \hline
\multicolumn{1}{|c|}{SQ2-states} & 1$^{+}\{A_2\},$2$^{+}\{B_2\},$2$^{-}\{E\}$
\\ \hline
\multicolumn{1}{|c|}{CH1-states} & 0$^{+}\{B_1\},$2$^{+}\{B_1\}$ \\ \hline
\multicolumn{1}{|c|}{CH2-states} & 1$^{-}\{B_2\},$1$^{-}\{A_2\}$ \\ \hline
\multicolumn{1}{|c|}{CH3-states} & 0$^{+}\{E\},$2$^{+}\{E\}$ \\ \hline
\multicolumn{1}{|c|}{D-S-D states} & 0$^{+}\{B_2\},$1$^{+}\{B_1\},$2$%
^{+}\{A_2\}$ \\ \hline
\multicolumn{1}{|c|}{T-S-T states} & 0$^{-}\{B_1\},$1$^{+}\{E\},$2$%
^{-}\{A_1\},$2$^{-}\{B_1\},$2$^{-}\{A_2\},$ \\ \hline
\multicolumn{1}{|c|}{} & 0$^{+}\{A_2\},$0$^{-}\{A_1\},$0$^{-}\{B_2\},$0$%
^{-}\{A_2\},$0$^{-}\{E\},$1$^{+}\{A_1\},$1$^{+}\{B_2\},$1$^{-}\{A_1\},$1$%
^{-}\{B_1\},\cdot \cdot \cdot $ \\ \hline
\end{tabular}

Table 2, A classification of the L$\leq 2$ states.

\hspace{1.0in}

\hspace{1.0in}

\begin{tabular}{|c|c|c|}
\hline
& type & energy \\ \hline
0$^{+}\{A_1\}$ & SQ1 & -0.5160 \\ \hline
0$^{+}\{B_1\}$ & CH1 & -0.4994 \\ \hline
0$^{+}\{E\}$ & CH3 & -0.3300 \\ \hline
0$^{+}\{B_2\}$ & D-S-D & -0.3145 \\ \hline
0$^{+}\{A_2\}$ &  & -0.3121 \\ \hline
\end{tabular}

Table 3, The energies of the 0$^{+}$ first-states from [6] in atomic unit.

\hspace{1.0in}

\hspace{1.0in}

\begin{tabular}{|c|c|c|c|}
\hline
& L=0 & L=1 & L=2 \\ \hline
A$_1$ & SQ, CH1 &  & CH1, D-S-D \\ \hline
B$_2$ & D-S-D & CH2 & SQ \\ \hline
B$_1$ & CH1, D-S-D &  & SQ, CH1 \\ \hline
A$_2$ & SQ & CH2 & D-S-D \\ \hline
E & CH2 & SQ, CH1 & CH2 \\ \hline
\end{tabular}

Table 4, The accessibility of regular shapes to the eigenstates of a
2-dimensional ($e^{+}e^{+}e^{-}e^{-}$) system. Refer to Table 1.

\hspace{1.0in}

\hspace{1.0in}

\begin{tabular}{|c|c|}
\hline
SQ1-states & 0\{A$_1$\},1\{$E$\},2\{B$_1$\} \\ \hline
SQ2-states & 0\{A$_2$\},2\{B$_2$\} \\ \hline
CH1-states & 0\{B$_1$\},2\{A$_1\}$ \\ \hline
CH2-states & 0\{$E$\},1\{B$_2$\},1\{A$_2$\},2\{$E$\} \\ \hline
D-S-D states & 0\{B$_2$\},2\{A$_2$\} \\ \hline
& 1\{A$_1$\},1\{B$_1$\} \\ \hline
\end{tabular}

Table 5, A classification of 2-dimensional L$\leq 2$ states. Refer to Table
2.

\end{document}